\documentclass{format170x240multiauthor}
\usepackage[T1]{fontenc}
\usepackage{makeidx}
   \makeindex
\usepackage{amsmath,format-thm}
\usepackage{textcomp}
\usepackage{graphicx}
\usepackage{cite}
\usepackage{url}
\usepackage{graphics}
\usepackage{fancybox}
\usepackage{latexsym,psfrag,epsf,epsfig,amssymb,rotating}
\usepackage{color}
\usepackage{multirow}
\usepackage{fullpage}
\usepackage{subfigure}
\usepackage{bm}

\usepackage{morefloats}
\usepackage{pifont}
\newcommand{\cmark}{\ding{51}}
\newcommand{\xmark}{\ding{55}}

\begin{document}

\chapter{Bayesian computational algorithms for social network analysis}
\label{BSNA}
\chapterauthor[Alberto Caimo, Isabella Gollini]
{Alberto Caimo\protect\footnote{Social Network Analysis Research Center, \\
                                                   Universit\`{a} della Svizzera italiana, Switzerland. \\
                                                   \tt{alberto.caimo@usi.ch}}, 
Isabella Gollini\protect\footnote{Department of Civil Engineering, \\
                                                   University of Bristol, United Kingdom. \\
                                                   \tt{isabella.gollini@bristol.ac.uk}}}

\section{Introduction}
\label{BSNA:intro}
Interest in statistical network analysis has grown massively in recent decades and its perspective and methods are now widely used in many scientific areas which involve the study of various types of networks for representing structure in many complex relational systems such as social relationships, information flows, protein interactions, etc.

Social network analysis is based on the study of social relations between actors so as to understand the formation of social structures by the analysis of basic local relations. Statistical models have started to play an increasingly important role because they give the possibility to explain the complexity of social behaviour and to investigate issues on how the global features of an observed network may be related to local network structures. The observed network is assumed to be generated by local social processes which depend on the self-organising dyadic relations between actors. The crucial challenge for statistical models in social network theory is to capture and describe the dependency giving rise to network global topology allowing inference about whether certain local structures are more common than expected.

Unfortunately the computational burden required to estimate these models is the main barrier to estimation. Recent theoretical developments and advances in approximate procedures have given the possibility to make important progress to overcome statistical inference problems. 

In this chapter we review some of the most recent computational advances in the rapidly expanding field of statistical social network analysis (see \cite{tow:whi:gol:mur12} for a recent review) using the \verb+R+ open-source software.

In particular we will focus on Bayesian estimation for two important families of models: exponential random graph models (ERGMs) and latent space models (LSMs) and we will provide the \verb+R+ code used to produce the results obtained in this chapter.

The chapter is organised as follows: in Section~\ref{BSNA:randomgraphs}, we introduce the basic notation for social network analysis. In Section~\ref{BSNA:statmodels}, we highlight the basic statistical work on social networks citing recent references to enable interested readers to learn more. In particular, our interest lies on describing exponential random graph models and latent space models.
In Section 1.4, we discuss Bayesian analysis for these two families of models and computational methods on a well-known dataset using the \verb+R+ software. Predictive goodness-of-fit diagnostics are also described at the end of the section.
We conclude in Section~\ref{BSNA:conclusions} with a discussion of some future challenges.

\section{Social networks as random graphs}
\label{BSNA:randomgraphs}
Networks are relational data that can be defined as a collection of nodes interacting with each other and connected in a pairwise fashion. In typical applications, the nodes represent a set actors of various kind (people, organisations, countries, etc.) and the set edges represent a specific relationship between them (friendship, collaboration, etc.).

From a statistical point of view, networks are relational data represented as mathematical graphs. A graph consists of a set of $n$ nodes and a set of $m$ edges which define some sort of relationships between pair of nodes called dyads.

The connectivity pattern of a graph can be described by an $n \times n$ adjacency matrix $y$ encoding the presence or absence of an edge between node $i$ and $j$:
\begin{equation*}
y_{ij} 
  =\left\{\begin{matrix}
          1, & \textrm{if }(i,j) \textrm{ are connected,}\\
          0, & \textrm{otherwise.}\\
          \end{matrix}
   \right.
\end{equation*}
Two nodes are adjacent or neighbours if there is an edge between them. If $y_{ij} = y_{ji}, \forall i, j$ then the adjacency matrix is symmetric and the graph is undirected, otherwise the graph is directed and it is often called digraph. Edges connecting a node to itself (self-loops) are generally not allowed in many applications and will not be considered in this context. The nature of the edges between nodes can take a range of values indicating the strength, frequency, intensity, etc. of the relation between a dyad. In this paper we consider only binary networks. According to the generally used notation, $y$ will be used to indicate both a random graph and its adjacency matrix.

\section{Statistical modelling approaches to social network analysis}
\label{BSNA:statmodels}
Many probability models have been proposed in order to summarise the connectivity structure of social networks by utilising their network statistics. 

The family of exponential random graph models (ERGMs) is a generalisation of various models which take different assumptions into account: the Bernoulli random graph model \cite{erd:ren59} in which edges are considered independent of each other; the $p_1$ model \cite{hol:lei81} where dyads are assumed independent, and its random effects variant the $p_2$ model \cite{van:sni:zij04}; and the Markov random graph model \cite{fra:str86} where each pair of edges is conditionally dependent given the rest of the graph. 
The family of latent space models has been proposed by \cite{hof:raf:han02} under the assumption that each node of the graph has a unknown position in a latent space and the probability of the edges are functions of those positions and node covariates. The latent position cluster model of \cite{han:raf:tan07} represents a further extension of this approach that takes account of clustering. 
Other latent variable modelling approaches are represented by stochastic blockmodels \cite{now:sni01} that involve block model structures whereby network nodes are partitioned into latent classes and the presence of any relationship between them depends on their block membership. 

\subsection{Exponential random graph models (ERGMs)}
\label{BSNA:ERGMs}
Introduced by \cite{hol:lei81} to model individual heterogeneity of nodes and reciprocity of their edges, the family of exponential random graph models (ERGMs) was generalised by \cite{fra:str86}, \cite{was:pat96} and \cite{sni:pat:rob:han06}. 
ERGMs constitute a broad class of network models (see \cite{rob:pat:kal:lus07} for an introduction) that assume that the topological structure of an observed
network $y$ can be explained in terms of the relative prevalence of a set of overlapping subgraph configurations $s(y)$ called network statistics:
\begin{equation}
p(y | \theta) = \frac{ \exp \{ \theta^t s(y)\} }
                             {z(\theta)}
\end{equation}
This equation states that the probability that an observed network $y$ given the set of parameters $\theta$ is equal to the exponent of an observed vector of network statistics $s(y)$ multiplied by its associated vector of unknown parameters $\theta$ divided by a normalising constant $z(\theta)$ to make all probabilities sum to one. The latter is calculated as the sum over all possible network configurations on the same set of $n$ nodes of the observed network. In practice $z(\theta)$ is computationally infeasible to calculate for non trivially-small networks.

\subsection{Latent space models (LSMs)}
\label{BSNA:LVMs}
Latent space models were introduced by \cite{hof:raf:han02} under the basic assumption that each node has an unknown position $z_i$ in a $d$-dimensional Euclidean latent space. Network edges are assumed to be conditionally independent given the latent positions, and the probability of an edge  between nodes is modelled as a function of their positions. 
Generally, in these models the smaller the distance between two nodes in the latent space, the greater their probability of being connected.
The likelihood function of  latent space models can be written as follows:

\begin{equation*}
p(y|z,\alpha)=\prod_{i\neq j}^N \frac{\exp(\alpha-d_{ij})^{y_{ij}}}{1+\exp(\alpha-d_{ij})}
\end{equation*}

The standard metric is the Euclidean distance (ED in Table~\ref{tab:pack}) and is defined as: $d_{ij} = |z_i- z_j|$. As an alternative the squared Euclidean distance (SED in Table~\ref{tab:pack}) is defined as: $d_{ij} = |z_i- z_j|^2$ and has been proposed by \cite{gol:mur15} for computational reasons (see~\ref{BSNA:BLVMs}). The latent positions are assumed to be Normally distributed, or having a Gaussian mixture model structure in case of the latent position cluster models (LPCMs), a generalisation of latent space models where latent clusters are assumed to be useful to explain data heterogeneity.
For strongly asymmetric graph, it is suggested to use the bilinear latent model setting $d_{ij} = z_i'z_j$ so that the probability of a link depends on the angle between two actors. This model is available in the \verb+latentnet+ package through the \verb+bilinear+ argument included in the \verb+ergmm+ function. 
All the presented latent space network models can be extended to incorporate covariate informations $x_{ij}$ introducing a parameter $\beta$, or the degree heterogeneity in sending or receiving links, these parameters are called sender and receiver if the network is directed, or sociality if the network is undirected \cite{Krivitsky2009}.

\section{Bayesian inference for social network models}
\label{BSNA:Bayes}
The Bayesian approach to statistical problems is probabilistic. Inference is based on the posterior distribution which is the conditional probability of the unknown quantities $\Omega$ given the data $y$. The posterior distribution extracts the information in the data and provide a complete summary of the uncertainty about the unknowns via Bayes' theorem:
\begin{equation}
p(\Omega | y) = \frac{p(y | \Omega) \; p(\Omega)}{p(y)}
\end{equation}
Bayesian analysis is able to give us a full probabilistic framework of uncertainty and this is something which is essential in the context of complex statistical modelling. Moreover recent research in social network analysis has demonstrated the advantages and effectiveness of probabilistic Bayesian approaches to relational data.
In this chapter we will focus on parameter inference so the uncertainties $\Omega$ will refer to the ERGM parameters $\theta$ or the LSM parameters $\alpha$ and latent positions $z$.

\subsection{R-based software tools}
\label{BSNA:Rtools}
Applied researchers interested in Bayesian statistics are increasingly attracted to \verb+R+ \cite{R} because of the ease of which one can code algorithms to sample from posterior distributions as well as the significant number of packages contributed to the Comprehensive R Archive Network (CRAN) that provide tools for Bayesian inference. 

\verb+R+ represents a useful tool for social network analysis with many advantages over traditional software packages. With a little coding and patience, one can produce ad hoc analyses and visualisations for the problem under study.
Moreover \verb+R+ has a huge set of statistical libraries so that end users can complement their social network analysis research with any analysis of your choosing within \verb+R+ environment.

In this section we briefly review Bayesian tools for ERGMs and LSMs:
\begin{itemize}
\item The \verb+Bergm+ package (version 3.0.1) \cite{cai:fri14} implements Bayesian analysis for ERGMs using the methods proposed by \cite{cai:fri11,cai:fri13,cai:mir15}. The package provides a comprehensive framework for Bayesian inference and model selection using Markov chain Monte Carlo (MCMC) algorithms.
\item The \verb+latentnet+ package (version 2.5.1) \cite{latentnet, latentnet2}, which is part of the \verb+statnet+ suite of packages \cite{han:hun:but:goo:mor07}, provides comprehensive toolsets for Bayesian analysis for latent position and cluster network models using MCMC procedures.
\item The \verb+VBLPCM+ package (version 2.4.3) \cite{sal:mur13} contains a collection of functions implementing variational Bayesian Inference for the latent position cluster model.
\item The \verb+lvm4net+ package (version 0.2) \cite{gol15} contains a collection of functions implementing fast variational Bayesian inference for latent space models.
\end{itemize}
Other \verb+R+ implementations of Bayesian methods for statistical social network models  include: \verb+RSiena+ \cite{rip:sni11} implementing stochastic actor-based models; \verb+hergm+ \cite{hergm} implementing hierarchical ERGMs with local dependence; \verb+sna+ (belonging to the \verb+statnet+ suite of packages) generating posterior samples from Butt's Bayesian network accuracy model using Gibbs sampling.

\section{Data}
\label{BSNA:data}
\begin{figure}[htp]
\centering
\includegraphics[scale=.8]{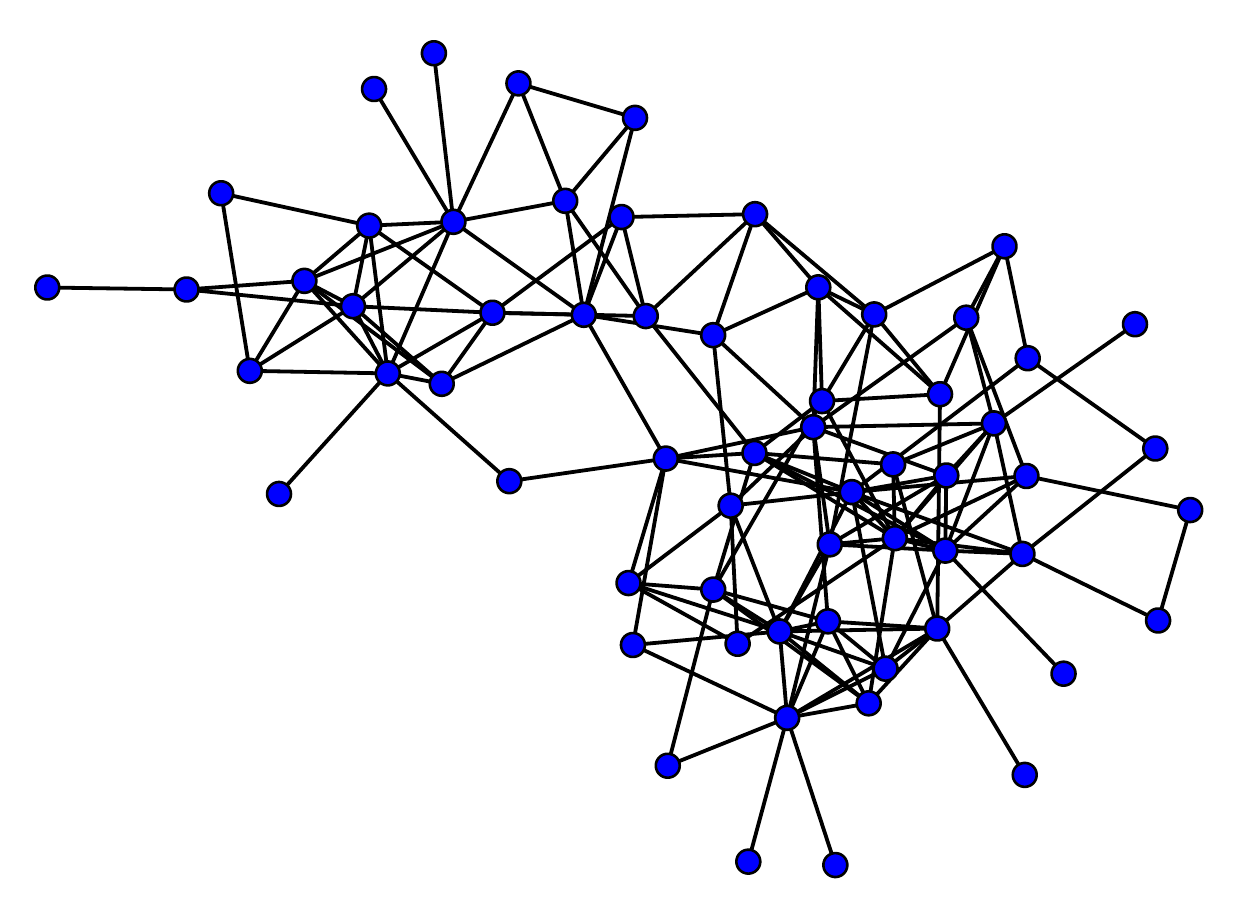}
\caption{Dolphin undirected network graph.}
\label{fig:Dolphins_graph}
\end{figure}

We demonstrate ideas and examples throughout the paper using the Dolphin network dataset, an undirected social network of frequent associations between 62 dolphins in a community living off Doubtful Sound, New Zealand (see Figure~\ref{fig:Dolphins_graph}), as compiled by \cite{lus:sch:boi:haa:slo:daw03}.
The results presented in this paper have been obtained using \verb+R+ version 3.1.3. To create, manipulate and visualise the observed network data $y$ we can use the function \verb+network+ and \verb+plot+ included in the \verb+statnet+ suite of packages.

\begin{verbatim}
y <- read.table("http://moreno.ss.uci.edu/dolphins.dat", 
                skip = 130)

y <- network(y, directed = FALSE)

plot(y, vertex.col = "blue")
\end{verbatim}

\subsection{Bayesian inference for exponential random graph models}
\label{BSNA:BERGMs}
Bayesian inference for ERGMs is based on the posterior distribution of $\theta$ given the data $y$:
\begin{equation}
p(\theta | y) 
= \frac{p(y | \theta) \; p(\theta)}{p(y)} 
= \frac{ \exp \{ \theta^t s(y)\} }
       {z(\theta)} \frac{p(\theta)}{p(y)}
\end{equation}

where $p(y)$ is the evidence or marginal likelihood of $y$ which is typically intractable. 

Standard MCMC methods such as the Metropolis-Hastings algorithm, can deal with posterior estimation as long as the target posterior density is known up to the model evidence $p(y)$. Unfortunately in the ERGM context the posterior density $p(\theta | y)$  of non-trivially small ERGMs includes two intractable normalising constants, the model evidence $p(y)$ and $z(\theta)$. For this reason, the ERGM posterior density is ``doubly intractable'' \cite{mur:gha:mac06}.

In order to carry out Bayesian inference for ERGMs, the \verb+Bergm+ package makes use of a combination of Bayesian algorithms and MCMC techniques \cite{cai:fri11,cai:fri13}. The approximate exchange algorithm circumvents the problem of computing the normalising constants of the ERGM likelihoods, while the use of multiple chains and efficient adaptive proposal strategies are able to speed up the computations and improve chain mixing quite significantly.

The approximate exchange algorithm implemented by the \verb+bergm+ function can be summarised in the following way:\\[.4cm]

For each chain, repeat until converge:\\[.3cm]
\begin{itemize}
\item[1)] generate $\theta'$ (using some proposal strategy)\\[.2cm] 
\item[2)] simulate $s(y')$ from ERGM likelihood (using standard MCMC procedures such as \cite{mor:han:hun08})\\[.2cm] 
\item[3)] update $\theta \rightarrow \theta'$ with the log of the probability:
\begin{equation*}
\min\left( 0,\; \left[ \theta - \theta'\right]^t \left[s(y') - s(y)\right]
  +\log\left[
   \frac{p(\theta')}
        {p(\theta)}\right]\right)
\end{equation*}
\end{itemize}

Let us consider the following three dimensional model including the number of edges and two new specification statistics e.g.: geometrically weighted edgewise shared partners (gwesp) and geometrically weighted non-edgewise shared partners (gwesp) \cite{hun:han06}:
\begin{center}
\begin{tabular}{ll}
\verb+gwnsp+ =& $e^{\phi_v} \sum_{k=1}^{n-2}
\left \{ 1-\left( 1 - e^{-\phi_v} \right)^{k} \right \} NEP_k(y)$  \\
\verb+gwesp+ =& $e^{\phi_v} \sum_{k=1}^{n-2}
\left \{ 1-\left( 1 - e^{-\phi_v} \right)^{k} \right \} EP_k(y)$
\end{tabular}
\end{center}
where the scale parameters $\phi_v = \phi_u=0.6$, and $EP_k(y)$ and $NEP_k(y)$ are respectively the number of connected and non-connected pairs of nodes with exactly $k$ common neighbours.

We can  use the \verb+bergm+ function to sample from the posterior distribution using the MCMC algorithm described above. In this example we use the parallel adaptive direction sampling (ADS) procedure described in \cite{cai:fri11} for step 1 and 1,200 iterations (\verb+main.iters+) for each chain. We set the number of MCMC chains to 9 by using the argument \verb+nchains+. The number of iterations used to simulate network statistics $s(y')$ at step 2 is defined by the argument \verb+aux.iters+ and it is set to $3,000$.

\begin{verbatim}
model <- y ~ edges + 
             gwnsp(.6, fixed = TRUE) + 
             gwesp(.6, fixed = TRUE)

post <- bergm(model,
              main.iters = 1200,
              aux.iters = 3000,
              nchains = 9)
              
bergm.output(post, lag = 200)
\end{verbatim}

\begin{figure}[htp]
\centering
\includegraphics[scale=.65]{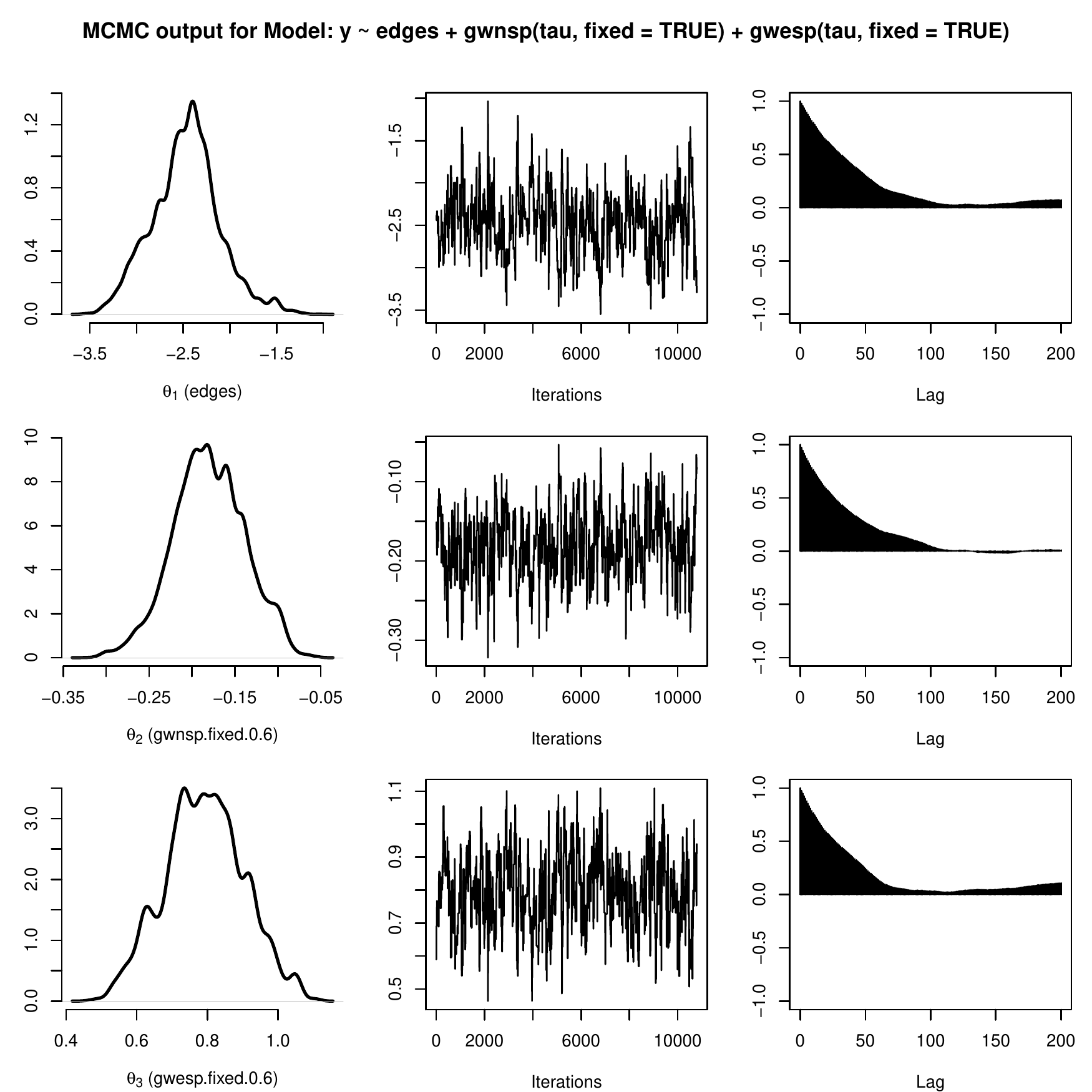}\\[.5cm]
\caption{MCMC diagnostics for the overall chain. The 3 plot columns are: estimated marginal posterior densities (left), traces (center) and autocorrelation plots (right).}
\label{fig:Bergm_post}
\end{figure}

The \verb+bergm.output+ function produces MCMC diagnostic plots (Figure~\ref{fig:Bergm_post}) and the estimated posterior means, standard deviations, and acceptance rates for each of the 9 chains and for the aggregated overall chain.
{\footnotesize
\begin{verbatim}
 MCMC results for Model: 
 y ~ edges + gwnsp(.6, fixed = TRUE) + gwesp(.6, fixed = TRUE) 

 Posterior mean: 
          theta1 (edges) theta2 (gwnsp.fixed.0.6) theta3 (gwesp.fixed.0.6)
Chain 1       -2.3512134               -0.1864153                0.7521076
Chain 2       -2.3889219               -0.1800818                0.7515701
Chain 3       -2.3362841               -0.1779192                0.7068975
Chain 4       -2.5628317               -0.1646549                0.7898211
Chain 5       -2.3709133               -0.1799828                0.7316151
Chain 6       -2.5407332               -0.1646283                0.7798916
Chain 7       -2.4301006               -0.1783869                0.7698418
Chain 8       -2.4523673               -0.1799679                0.8017089
Chain 9       -2.3681535               -0.1789971                0.7424549

 Posterior sd: 
          theta1 (edges) theta2 (gwnsp.fixed.0.6) theta3 (gwesp.fixed.0.6)
Chain 1       0.33244522               0.03951240               0.11308154
Chain 2       0.43199257               0.04669447               0.11585996
Chain 3       0.37344505               0.04083082               0.10625747
Chain 4       0.41110025               0.04962470               0.11782669
Chain 5       0.48867437               0.05371030               0.14904932
Chain 6       0.36796911               0.04055858               0.13496058
Chain 7       0.42739511               0.04311092               0.14186529
Chain 8       0.48943818               0.05430663               0.12666345
Chain 9       0.38717484               0.04531716               0.12718701

          Acceptance rate:
Chain 1          0.1316667
Chain 2          0.1375000
Chain 3          0.1158333
Chain 4          0.1550000
Chain 5          0.1475000
Chain 6          0.1566667
Chain 7          0.1700000
Chain 8          0.1525000
Chain 9          0.1500000


 Overall posterior density estimate: 
           theta1 (edges) theta2 (gwnsp.fixed.0.6) theta3 (gwesp.fixed.0.6)
Post. mean     -2.4223910              -0.17678157                0.7584343
Post. sd        0.4222507               0.04675703                0.1296237

 Overall acceptance rate: 0.15
\end{verbatim}
}

In this example, we can observe a low density effect expressed by the negative value of the posterior mean of the edge effect parameter ($\theta_1$) combined with the negative value of multiple connectivity ($\theta_2$) and positive value of transitivity parameter ($\theta_3$).

\subsection{Bayesian inference for latent space models}
\label{BSNA:BLVMs}

A fully Bayesian approach for latent space models allows the estimation of all the parameters and latent positions simultaneously e.g. via MCMC sampling or variational approximation.

In this paragraph, we perform an empirical Bayesian analysis in order to compare different computational approaches for the visualisation and prediction properties of LSMs with and without clustering. To carry out this type of analysis we can use the following \verb+R+ packages: \verb+latentnet+ \cite{latentnet}, \verb+VBLPCM+ \cite{sal:mur13} and \verb+lvm4net+ \cite{gol15}. Their main features of these packages are shown in Table~\ref{tab:pack}.

\begin{table}[ht]
\centering
\caption{Comparison of the main features of the packages for latent space modeling}\label{tab:pack}
\begin{tabular}{|l|cc|cc|c|}
  \hline
 & \multicolumn{2}{c|}{\textbf{Model}} &\multicolumn{2}{c|}{\textbf{Inference Method}} & \textbf{Clustering}\\
& ED & SED & MCMC & Variational & \\ 
  \hline
\verb+latentnet+ & \cmark & \xmark & \cmark & \xmark & \cmark \\ 
\verb+VBLPCM+ & \cmark & \xmark & \xmark & \cmark & \cmark \\ 
\verb+lvm4net+ & \xmark & \cmark & \xmark & \cmark & \xmark \\
   \hline
\end{tabular}
\end{table}



The \verb+latentnet+ package uses Bayesian MCMC algorithms so the model estimation is computationally very expensive, and the times to estimate the model can become extremely large even for networks of the order of hundreds of nodes. For this reason the variational Bayes approach to estimate the latent space model and the latent position cluster model in order to make feasible the modelling of larger networks \cite{sal:mur13,gol:mur15}.
The basic idea of this method is to find a lower bound of the log-likelihood by introducing a variational posterior distribution $q$ and maximize it \cite{jor:gha:jaa:sau99}. 
The posterior probability of the unknown parameters $(z,\alpha)$ can be written in the following form:
\begin{equation*}
p(z,\alpha|y)=Cp(y|z,\alpha)p(\alpha)\prod_{i=1}^Np(z_i),
\end{equation*}
where $C$ is the unknown normalising constant. In the \verb+VBLPCM+ package, a hierarchical prior structure is taken into consideration.

In \cite{gol:mur15}, the variational posterior $q(z,\alpha|y)$ is defined in the following way:
\begin{equation*}\label{q.var}
q(z,\alpha|y)=q(\alpha)\prod_{i=1}^Nq(z_i),
\end{equation*}
where $ q(\alpha)=\mathcal{N}(\tilde{\xi},\tilde{\psi}^2)$ and $q(z_i)=\mathcal{N}(\tilde{z_i},\tilde{\Sigma})$.

The idea of using the squared Euclidean distance in the LSM was proposed by \cite{gol:mur15} in order to have less approximation to be made in the variational estimation procedure.

In the \verb+latentnet+ package, we use the function called \verb+ergmm+ to estimate the posterior distribution of the LSM parameters and latent positions. The argument \verb+d+ refers to the dimension of the latent space, which we set equal to 2 to make the visualisation of the latent positions of the nodes easier.

\begin{verbatim}
  post.latentnet <- ergmm(y ~ euclidean(d = 2))
\end{verbatim}

In the \verb+VBLPCM+ package, we can use the function called \verb+vblpcmfit+ to estimate the posterior distribution of the LPCM by specifying the number of clusters. In order to estimate a LSM we consider one cluster by setting the argument \verb+G+ equal to 1.

\begin{verbatim}
  post.vblpcm <- vblpcmfit(vblpcmstart(y, G = 1, d = 2))
\end{verbatim}

It is important to notice that the variational maximisation algorithm is subject of the risk of reaching local maximum. 

The package \verb+VBLPCM+ provides a special function called \verb+vblpcmstart+ to generate initial latent positions. This algorithm is based on the Fruchterman-Reingold method by default (argument \verb+START+), but there is also the possibility of using random values, geodesic distances or Graph Laplacian methods. In this function other model features such as sociality effects, and node covariates can also be specified. 

In \verb+lvm4net+ we use the function called \verb+lsm+ to estimate the posterior distribution of the LSM parameters and latent positions using a variational inferential approach. This function makes use of the Fruchterman-Reingold method to set the initial positions by default. Multi-start procedure can be implemented by changing the value associated to the argument \verb+nstart+ and only the values reaching the maximum are stored. It is also possible to start from random initial positions by setting the argument \verb+randomZ+ equal to \verb+TRUE+. 

From Table~\ref{tab:compare}, we can see that the \verb+lsm+ function is much faster than the \verb+ergmm+ function. In this case, the squared Euclidean distance is used.

\begin{verbatim}
post.lvm4net <- lsm(y[,], D = 2) 
\end{verbatim}

\begin{table}[ht]
\centering
\caption{Timings in seconds to fit LSMs (no clustering, G = 1).}\label{tab:compare}
\begin{tabular}{l|c}
  \hline
& Time in sec. \\ 
  \hline
\verb+latentnet+ & 111.20 \\ 
\verb+VBLPCM+ & 14.02 \\ 
\verb+lvm4net+ & 6.47 \\ 
   \hline
\end{tabular}
\end{table}

The latent positions are invariant under rotation, reflection and translations. For this reason we can match the rotations using the \verb+rotXtoY+ function (included in \verb+lvm4net+) in order to visualise and compare the latent positions estimated by the three approaches using the \verb+plotY+ function (included in \verb+lvm4net+).

\begin{verbatim}
Z <- post.lvm4net$lsmEZ
Zm <- rotXtoY(post.latentnet$mkl$Z,Z)$X
Zv <- rotXtoY(post.vblpcm$V_z,Z)$X

plotY(y[,], EZ = Zm, main = "latentnet")
plotY(y[,], EZ = Zv, main = "VBLPCM")
plotY(y[,], EZ = Z, main = "lvm4net")
\end{verbatim}

\begin{figure}[htp]
\centering
\includegraphics[scale=.85]{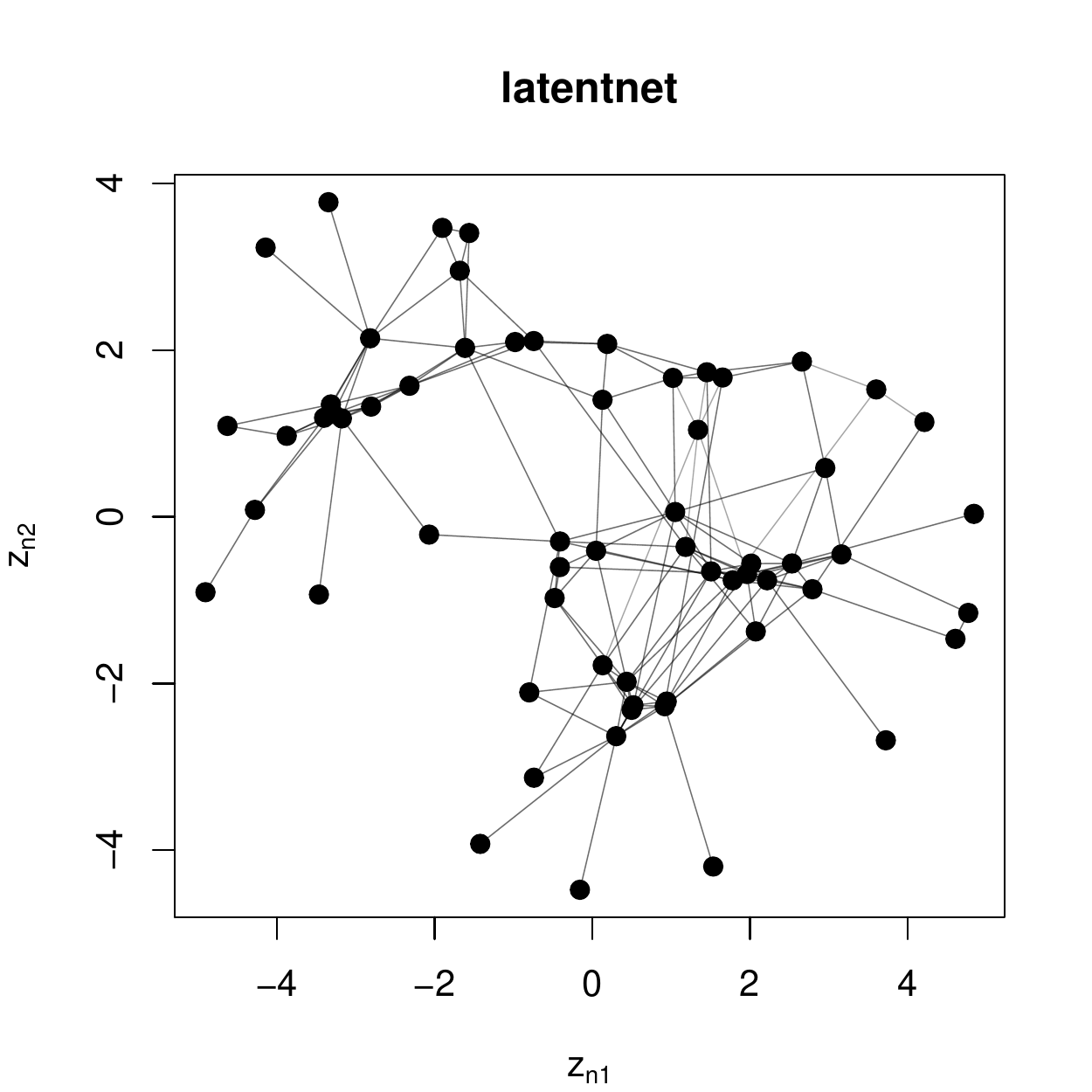}
\caption{Latent positions obtained by using the \texttt{latentnet} package.}
\label{fig:latentnet_plot}
\end{figure}

\begin{figure}[htp]
\centering
\includegraphics[scale=.85]{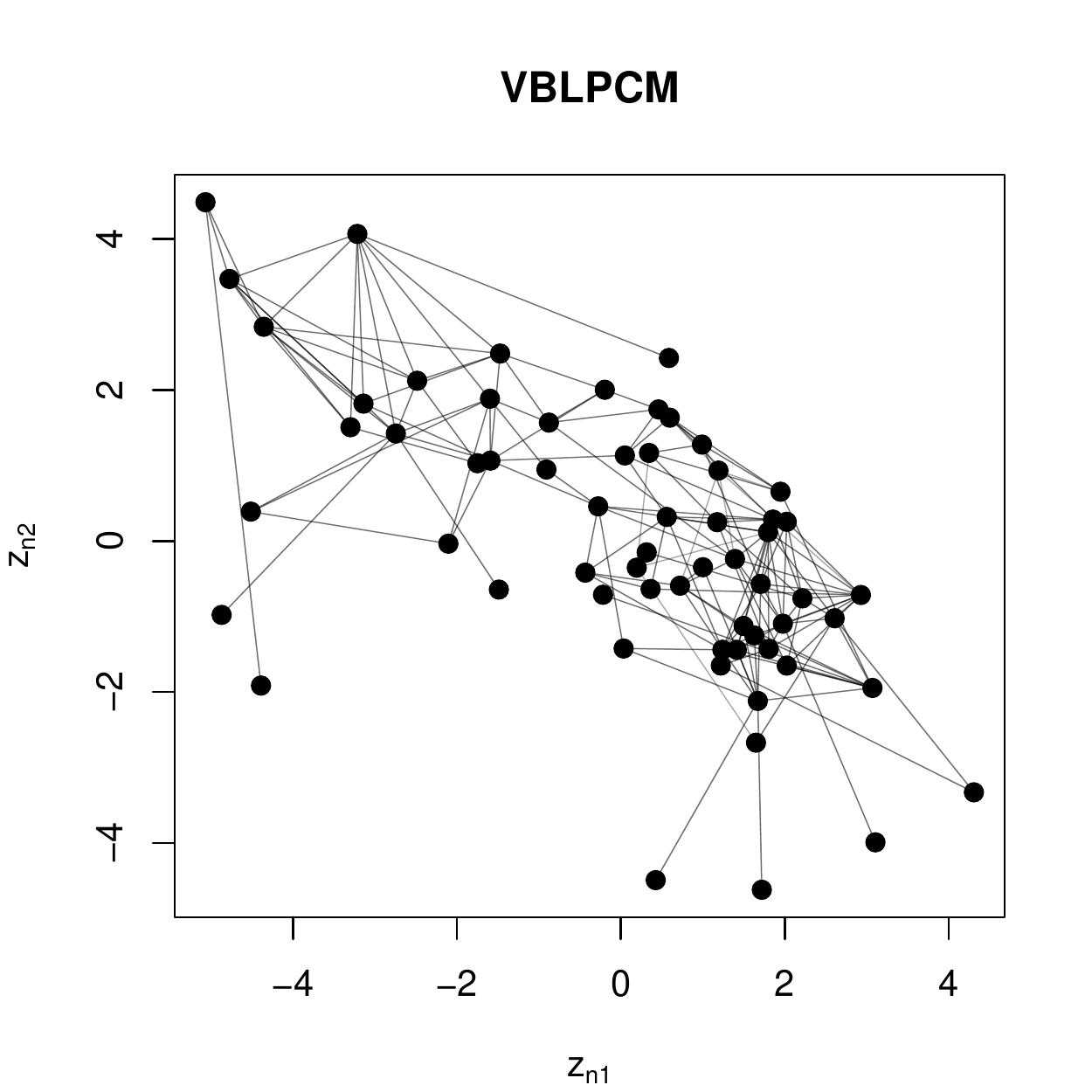}
\caption{Latent positions obtained by using the \texttt{VBLPCM} package.}
\label{fig:vblpcm_plot}
\end{figure}

\begin{figure}[htp]
\centering
\includegraphics[scale=.85]{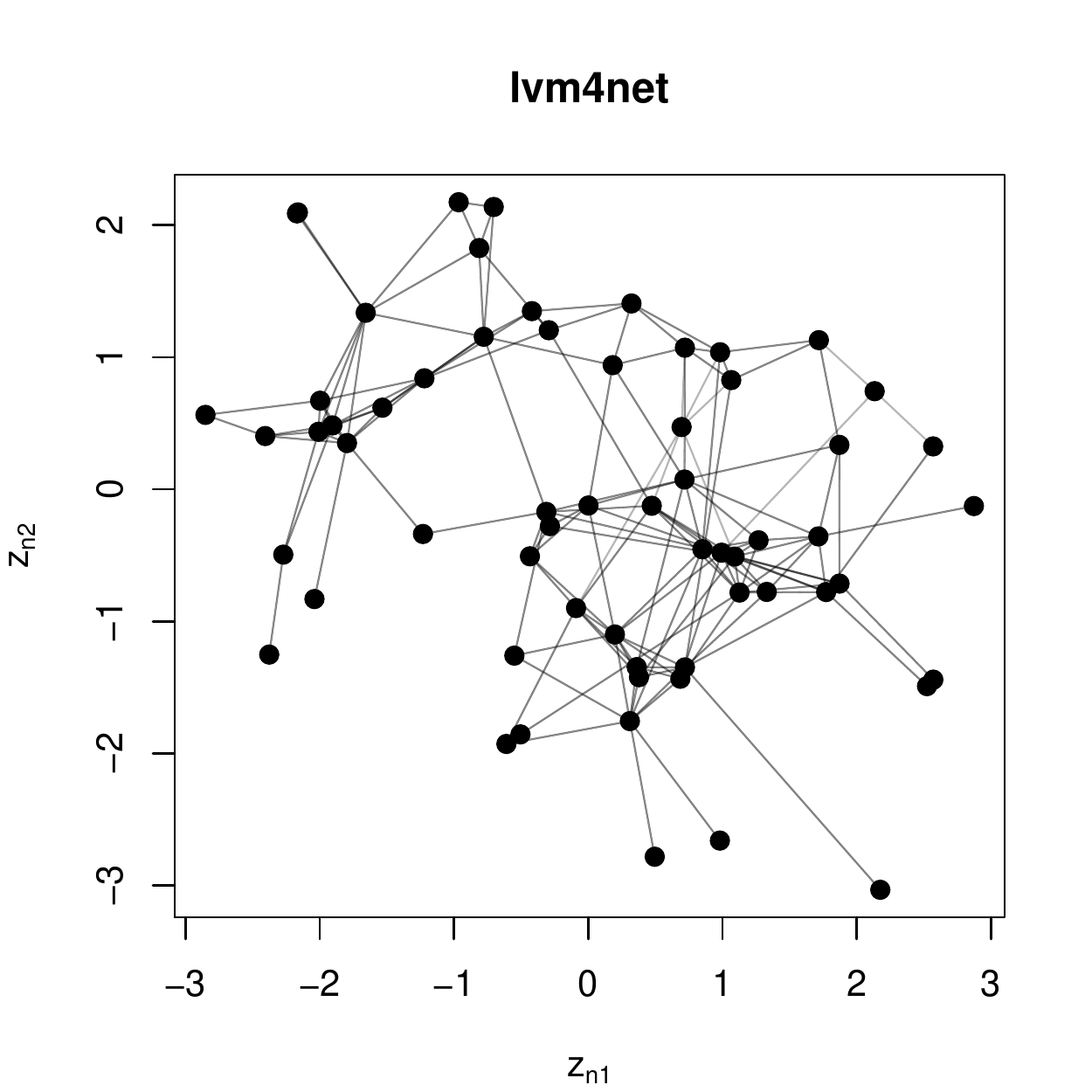}
\caption{Latent positions obtained by using the \texttt{lvm4net} package.}
\label{fig:lvm4net_plot}
\end{figure}

In Figures~\ref{fig:latentnet_plot}, \ref{fig:vblpcm_plot}, \ref{fig:lvm4net_plot} we can see the estimated latent positions obtained using the three packages. In this example, the visualisation results obtained from \verb+latentnet+ and \verb+lvm4net+ are similar even though the distance model adopted is different.

Latent position cluster models (LPCMs) are latent space models which incorporate a Gaussian mixture model structure for the latent positions of nodes in the latent space in order to accommodate the clustering of nodes in the network. The \verb+latentnet+ and \verb+VBLPCM+ packages can be used to estimate latent position cluster models by fixing the number of clusters (\verb+G+). For our toy example, we choose 2 clusters.

\begin{verbatim}
post.latentnet.G2 <- ergmm(y ~ euclidean(d = 2, G = 2))
\end{verbatim}

\begin{figure}[htp]
\centering
\includegraphics[scale=.7]{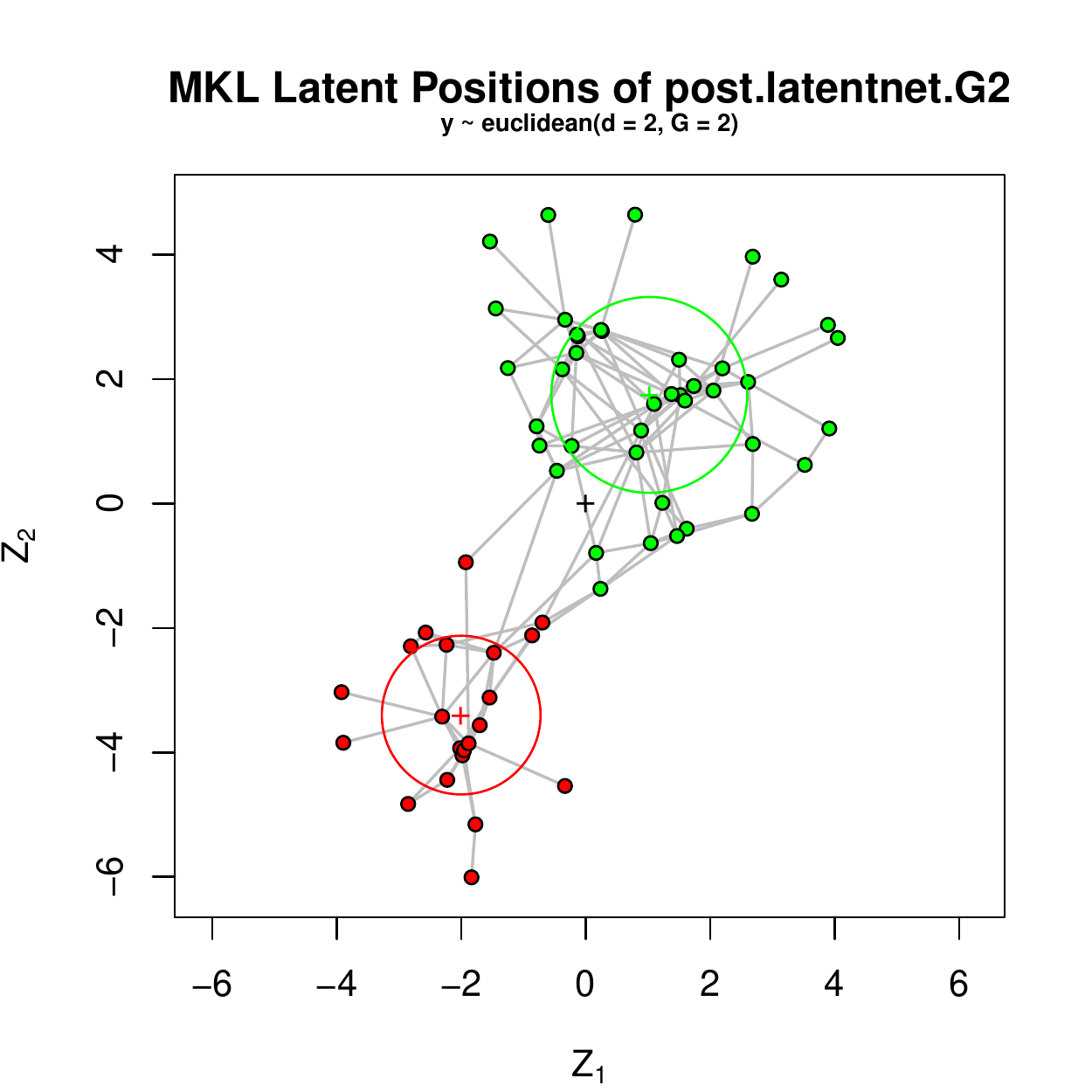}
\caption{Estimated latent positions from LPCM with $2$ clusters obtained by using the \texttt{latentnet} package.}
\label{fig:latentnet2_plot}
\end{figure}

\begin{verbatim}
post.vblpcm.G2 <- vblpcmfit(vblpcmstart(y, G = 2, d = 2))
\end{verbatim}

\begin{figure}[htp]
\centering
\includegraphics[scale=.7]{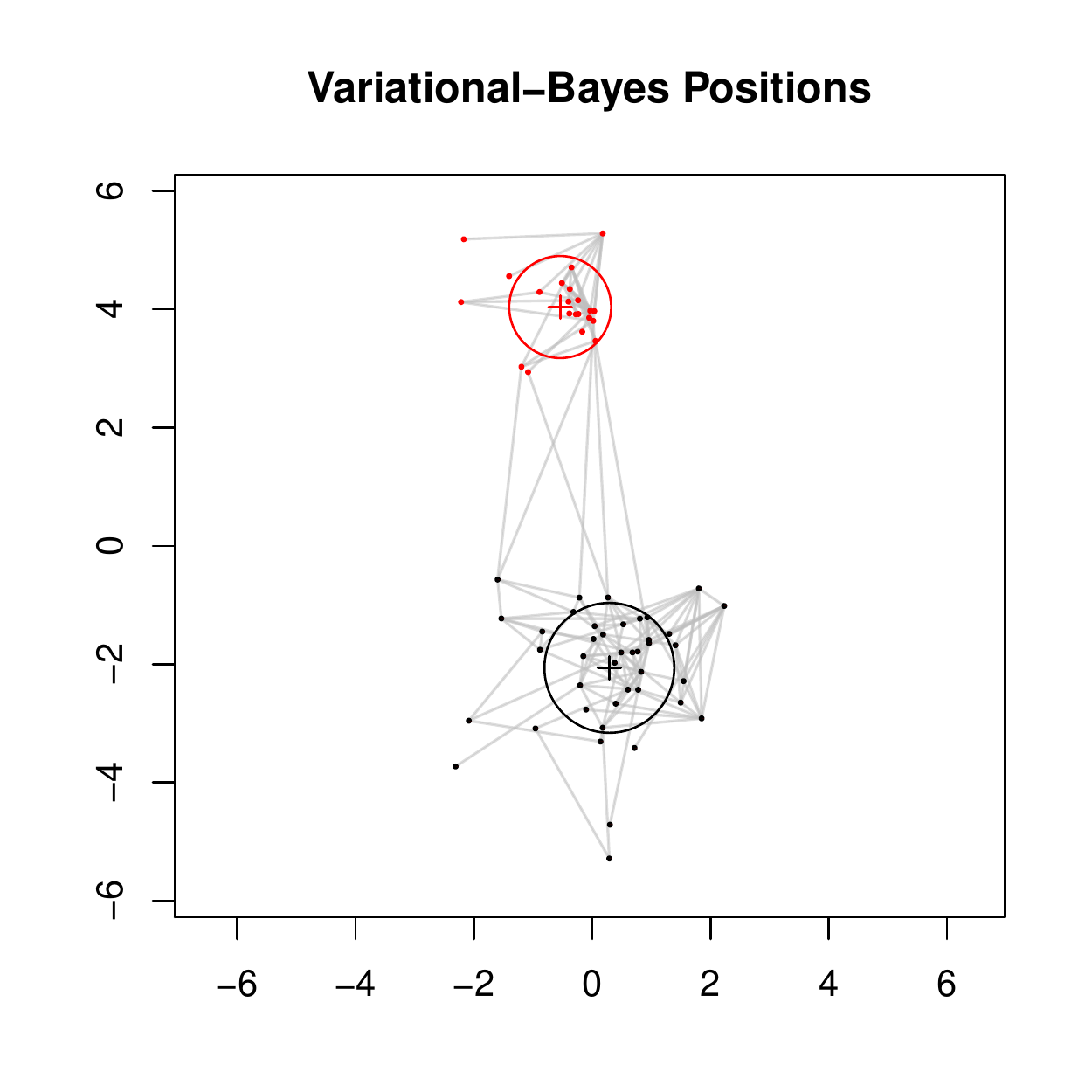}
\caption{Estimated latent positions from LPCM with $2$ clusters obtained by using the \texttt{VBLPCM} package.}
\label{fig:vblpcm2_plot}
\end{figure}

In Figures~\ref{fig:latentnet2_plot} and \ref{fig:vblpcm2_plot} we can see the latent positions and the clusters returned by the two packages. The two algorithms give very similar results as they find latent groups differing of just one node.

\begin{table}[ht]
\centering
\caption{Timings in seconds to fit LPCM with two clusters (G = 2).}\label{tab:compare2}
\begin{tabular}{l|c}
  \hline
& Time in sec. \\ 
  \hline
\verb+latentnet+ & 86.27 \\ 
\verb+VBLPCM+ & 11.95 \\
   \hline
\end{tabular}
\end{table}

From Table~\ref{tab:compare2} it is possible to notice that the \verb+VBLPCM+ package is much faster than the \verb+latentnet+ package. 

The \verb+latentnet+ package gives exact estimates as they are based on MCMC simulations from the posterior distribution. However it only allows to deal with small networks whereas the approximate approaches of the \verb+VBLPCM+ and \verb+lvm4net+ packages are able to handle networks on thousands of nodes.

\subsection{Predictive goodness-of-fit (GoF) diagnostics}
\label{BSNA:GOF}
An important feature of the Bayesian approach is to make available procedures to establish whether the estimated parameter posterior of the model achieves a good fit to the key topological features of the observed network.

The \verb++ function included in the \verb+Bergm+ package provides a useful tool for assessing Bayesian goodness-of-fit so as to examine the fit of the data to the posterior model obtained by the \verb+bergm+ function. The observed network data are compared with a set of networks simulated from independent parameter values of the posterior density estimate. This comparison is made in terms of high-level network statistics not explicitly included in the model \cite{hun:goo:han08}.

\begin{figure}[htp]
\centering
\includegraphics[scale=.8]{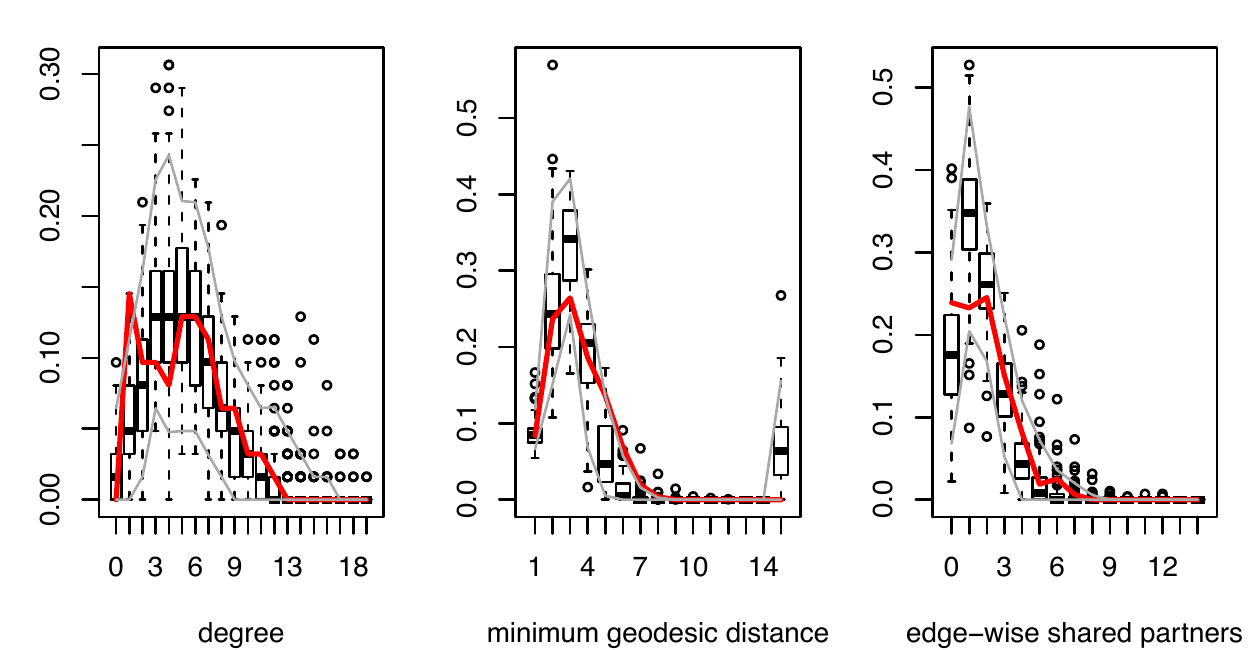}\\[.5cm]
\caption{GoF diagnostics for ERGM (\texttt{Bergm} package): The red line displays the goodness of fit statistics for the observed data together
with boxplots of GoF network statistics based on 100 simulated networks from the posterior distribution.}
\label{fig:Bergm_bgof}
\end{figure}

The \verb+R+ code below is used to compare some high level network statistics observed in the Dolphin network with a series of network statistics simulated from $100$ random realisations of the estimated posterior distribution \verb+post.est+ using $10,000$ auxiliary iterations for the network simulation step. The \verb+bgof+ function included in the \verb+Bergm+ package returns the plots shown in Figure~\ref{fig:Bergm_bgof}.

\begin{verbatim}
bgof(post, 
     n.deg = 20,
     n.dist = 15,
     n.esp = 15)
\end{verbatim}

In Figure~\ref{fig:Bergm_bgof} we see, based on the various GoF statistics, that the networks simulated from the posterior distribution are in reasonable agreement with the observed network. We can therefore conclude that the model is a reasonable fit to the data, despite its simplicity.

\begin{figure}[htp]
\centering
\includegraphics[scale=.8]{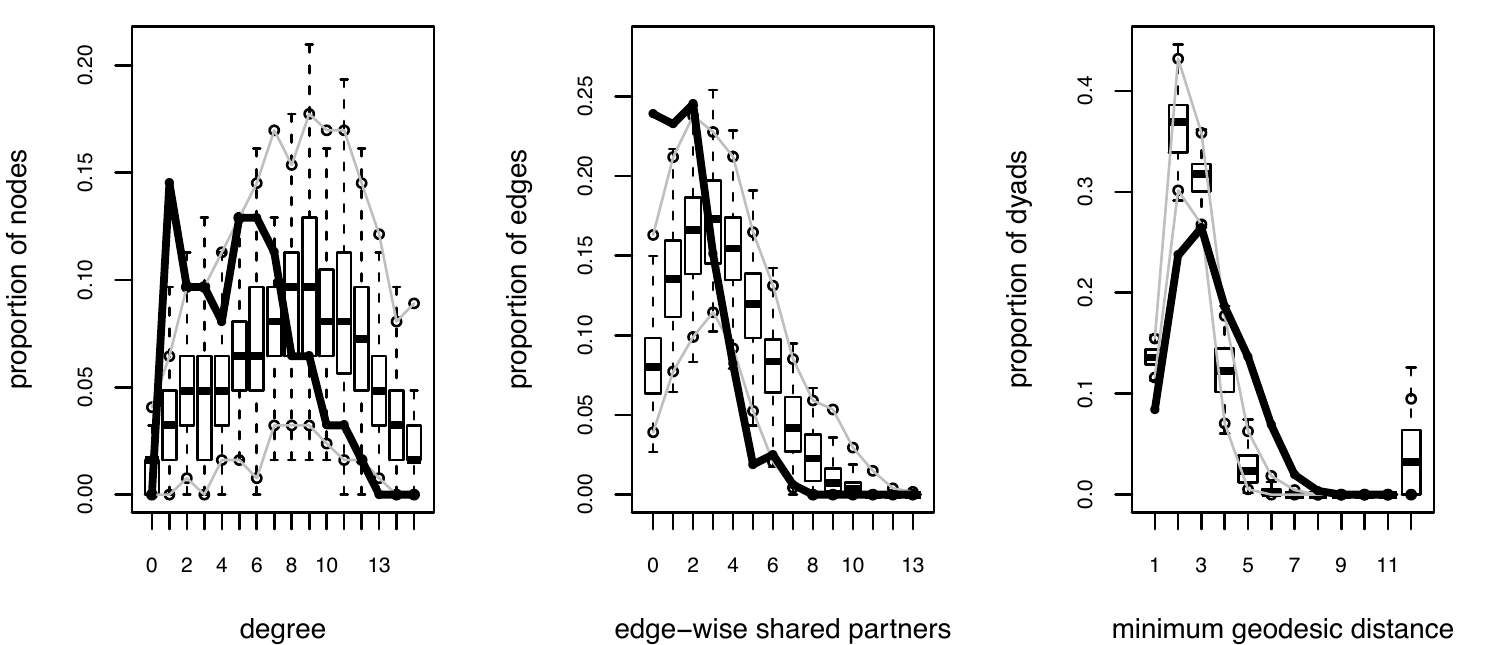}\\[.5cm]
\caption{GoF diagnostics for LSM (\texttt{latentnet} package): The solid black line displays the goodness of fit statistics for the observed data together with boxplots of GoF network statistics based on 100 simulated networks from the posterior distribution.}
\label{fig:gf_latentnet}
\end{figure}

\begin{figure}[htp]
\centering
\includegraphics[scale=.8]{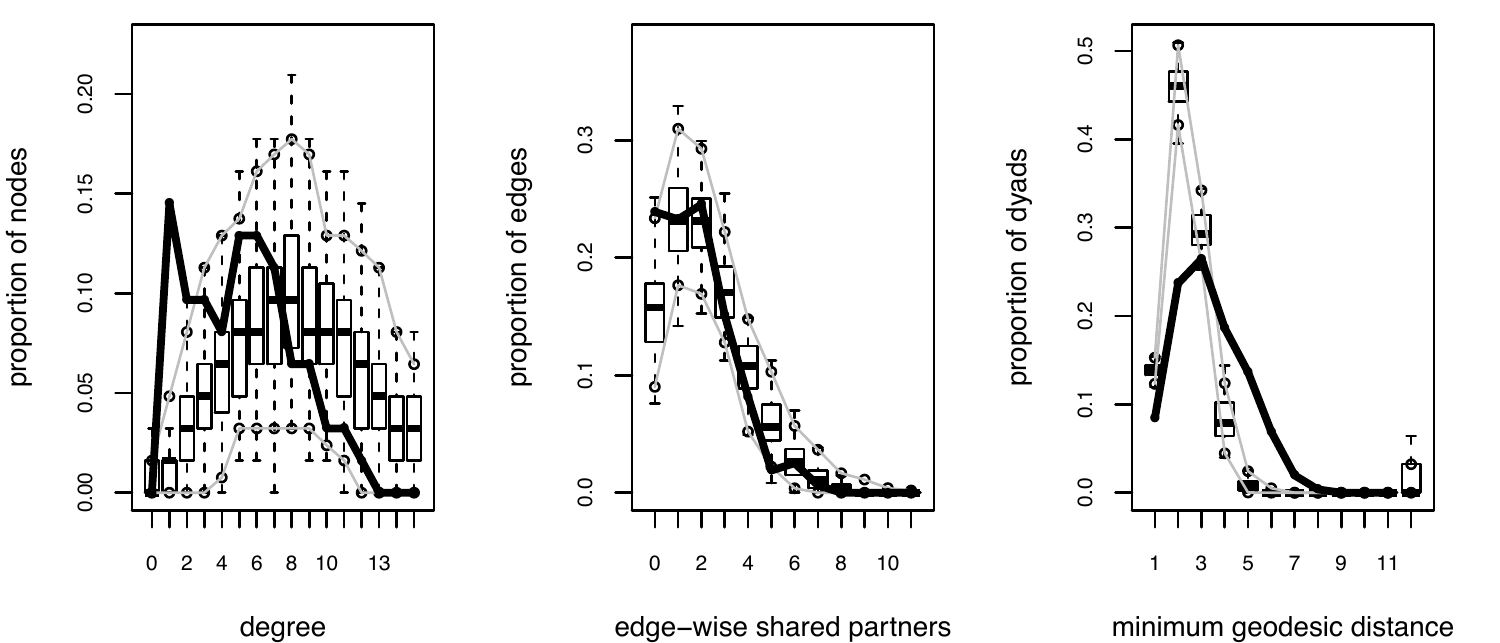}\\[.5cm]
\caption{GoF diagnostics for LSM (\texttt{VBLPCM} package): The solid black line displays the goodness of fit statistics for the observed data together with boxplots of GoF network statistics based on 100 simulated networks from the posterior distribution.}
\label{fig:gf_vblpcm}
\end{figure}

\begin{figure}[htp]
\centering
\includegraphics[scale=.8]{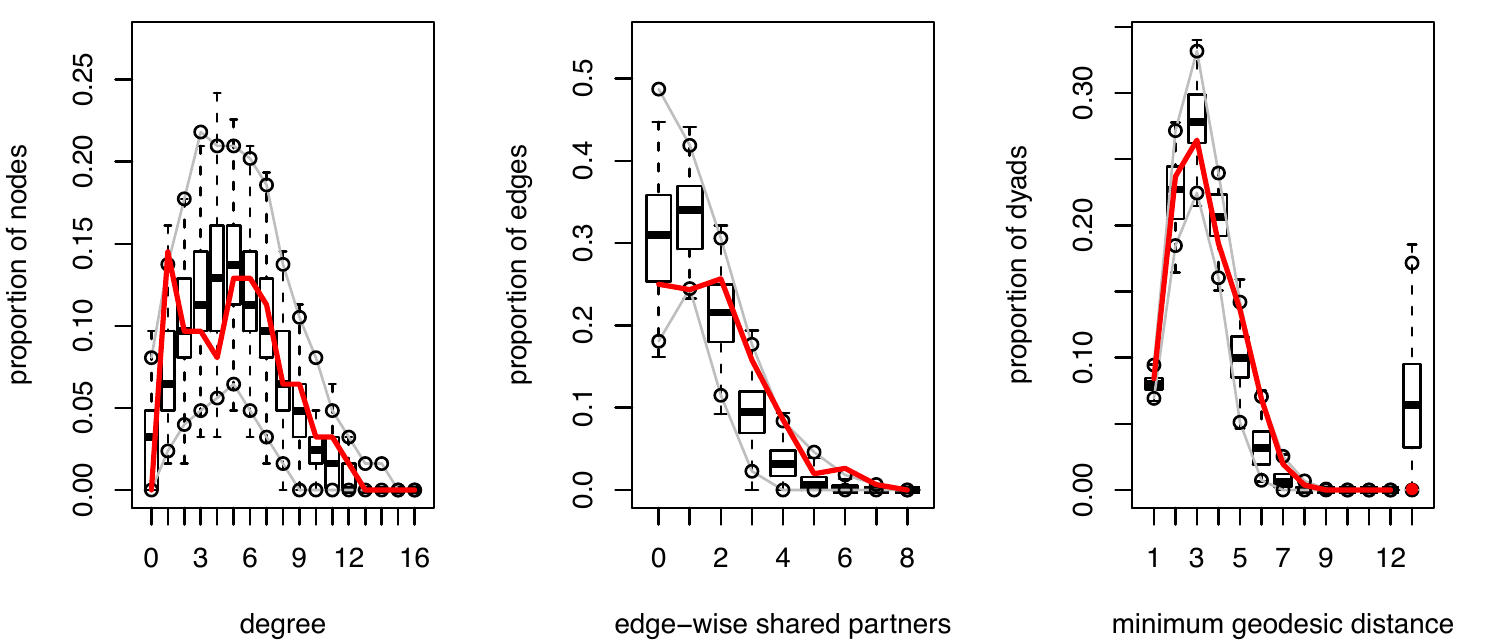}\\[.5cm]
\caption{GoF diagnostics for LSM (\texttt{lvm4net} package): The red line displays the goodness of fit statistics for the observed data together
with boxplots of GoF network statistics based on 100 simulated networks from the posterior distribution.}
\label{fig:gf_lvm4net}
\end{figure}

In the LSM context, it is possible to use the \verb+gof+ function included in \verb+latentnet+ and \verb+VBLPCM+ and the \verb+goflsm+ function included in the \verb+lvm4net+ package to perform posterior GoF diagnostics. The \verb+GOF+ argument can be used to set the types of GoF statistics we want to analyse. Figures~\ref{fig:gf_latentnet}, \ref{fig:gf_vblpcm}, and \ref{fig:gf_lvm4net} display the GoF plots.

\begin{verbatim}
gf.latentnet <- gof(post.latentnet, 
                    GOF = ~ degree + esp + distance)
plot(gf.latentnet)

gf.vblpcm <- gof(post.vblpcm, 
                 GOF = ~ degree + esp + distance)
plot(gf.vblpcm)

gf.lvm4net <- goflsm(post.lvm4net, 
                     Y = y[,], 
                     stats = c("degree", "esp", "distance"), 
                     doplot = FALSE)
plot(gf.lvm4net)
\end{verbatim}

The GoF analysis indicates that the LSM estimated by using the variational approximation with squared Euclidean distance implemented in the \verb+lvm4net+ package displays a better fit of the model to the data compared to the other two approaches.

To display the GoF diagnostics for the LPCMs estimated above, we can use the same \verb+R+ functions.

\begin{verbatim}
gf.latentnet.G2 <- gof(post.latentnet.G2, 
                    GOF = ~ degree + esp + distance)

plot(gf.latentnet.G2)

gf.vblpcm.G2 <- gof(post.vblpcm.G2, 
                 GOF = ~ degree + esp + distance)

plot(gf.vblpcm.G2)
\end{verbatim}

\begin{figure}[htp]
\centering
\includegraphics[scale=.8]{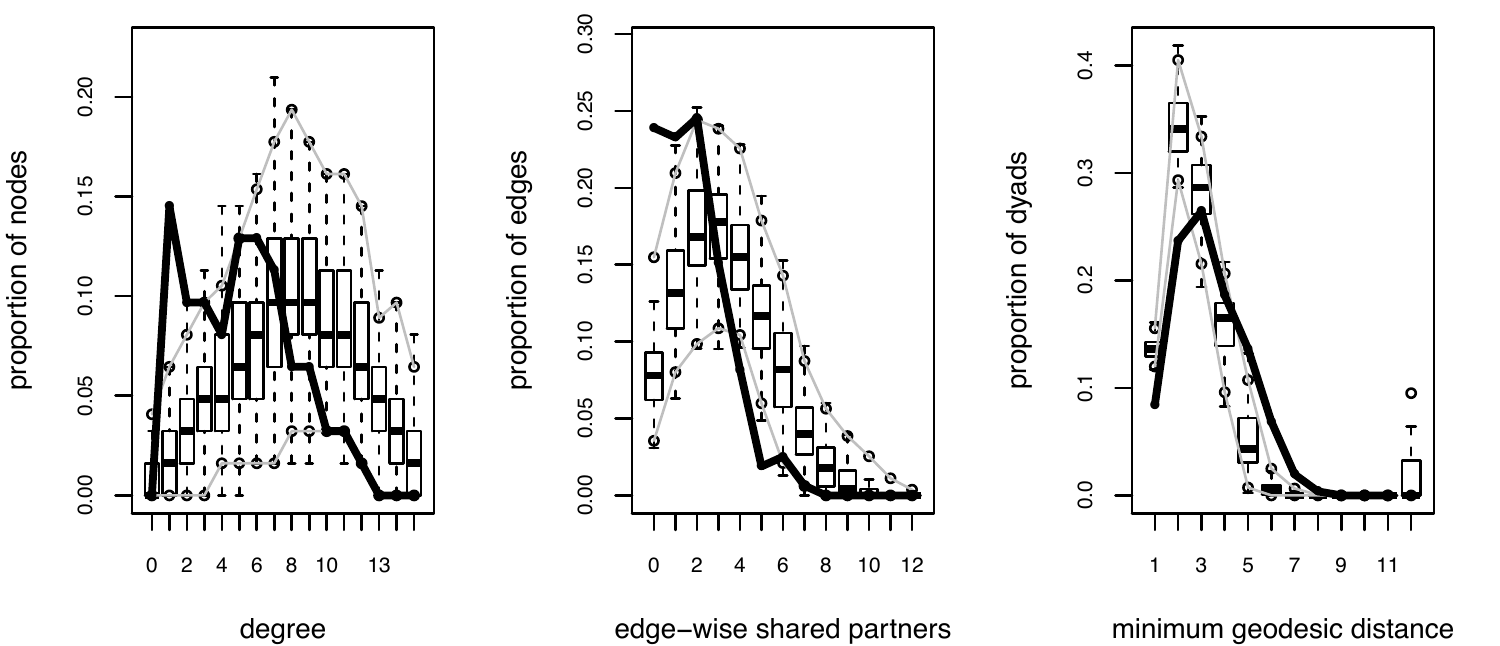}\\[.5cm]
\caption{GoF diagnostics for LPCM with 2 clusters (\texttt{latentnet} package): The solid black line displays the goodness of fit statistics for the observed data together with boxplots of GoF network statistics based on 100 simulated networks from the posterior distribution.}
\label{fig:gf_latentnet2}
\end{figure}

\begin{figure}[htp]
\centering
\includegraphics[scale=.8]{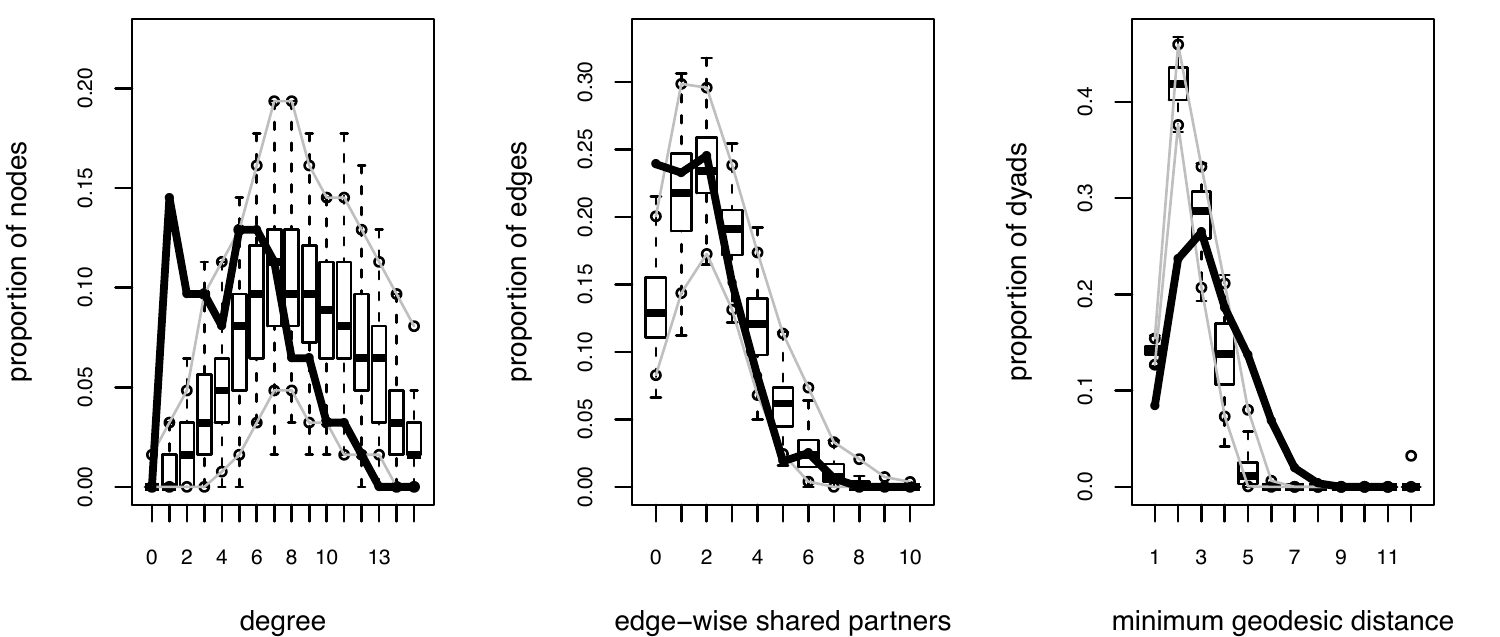}\\[.5cm]
\caption{GoF diagnostics for LPCM with 2 clusters (\texttt{VBLPCM} package): The solid black line displays the goodness of fit statistics for the observed data together with boxplots of GoF network statistics based on 100 simulated networks from the posterior distribution.}
\label{fig:gf_vblpcm2}
\end{figure}

From Figures~\ref{fig:gf_latentnet2} and \ref{fig:gf_vblpcm2} we can see that the \verb+VBLPCM+ package has a better fit to the data in terms of edgewise shared partners distributions compared to the \verb+latetnet+ package. For this example, the inclusion of 2 clusters does not seem to produce a significant  improvement in terms of GoF with respect to the LSM without clustering.

\section{Conclusions}
\label{BSNA:conclusions}
This chapter provided an overview of a number of social network models emphasising the computational perspective. In fact, the most important issue associated to statistical social network models is concerting their computational complexity which requires the development of efficient inferential algorithms and software able to deal with the increasing size of relational data available.

In particular, we have presented some recent advanced Bayesian approaches to parameter estimation of exponential random graph models and latent variable network models. We demonstrated that Bayesian inference is a very helpful and powerful approach allowing a formal treatment of uncertainty using the rules of probability. 

We discussed how Bayesian parameter estimation for exponential random graph models and latent space models is a computationally intensive problem that can be tackled using advanced MCMC and variational techniques. We illustrated the main capabilities of the \verb+Bergm+, \verb+latentnet+, \verb+VBLPCM+ and \verb+lvm4net+ packages for the open-source \verb+R+ software through a tutorial analysis of a well-known social network dataset.
For each modelling approach we have also considered a Bayesian graphical test of goodness of fit to assess whether or not a given parametric model is compatible with the observed network data. 

Advances in the Bayesian methodology and computing will prove crucial to effectively capture heterogeneity and organise different sources of information commonly available in social network data. For this reason, we believe statistical social network analysis will became fertile ground for interdisciplinary research in advanced statistics and social network analysis applications.

\newpage
\bibliography{BIBLIOGRAPHY}

\end{document}